\documentclass[aps,prd,twocolumn,showpacs,twoside]{revtex4}
\usepackage{graphicx,amsmath,bm}

\begin{document}

\newcommand{\tlab}{T_{\mbox{\scriptsize lab}}}
\newcommand{\plab}{p_{\mbox{\scriptsize lab}}}
\newcommand{\pppi}{\bar p p \rightarrow \pi^-\pi^+}
\newcommand{\ppkk}{\bar p p \rightarrow K^-K^+}
\newcommand{\s}{^3\!S_1}
\newcommand{\p}{^3\!P_0}
\newcommand{\ana}{A_{0n}}
\newcommand{\dsig}{d\sigma/d\Omega}
\newcommand{\pvec}{{\mathbf p}}
\newcommand{\rvec}{{\mathbf r}}
\newcommand{\D}{{\mathcal D}^s_{\lambda \lambda'}}
\newcommand{\wick}{\theta_{\mathrm{w}}}
\newcommand{\cm}{{\mathrm{cm}}}
\newcommand{\spi}{S_{\pi}}
\newcommand{\dvec}{\mathrm{d}{\mathbf p}}

\title{A relativistic treatment of pion wave functions in the annihilation $\pppi$}
\author{B.~El-Bennich}
\email{bennich@physics.rutgers.edu}
\author{W.M.~Kloet}
\affiliation{Department of Physics and Astronomy, Rutgers University, \\ 
             136 Frelinghuysen Road, Piscataway, New Jersey 08854-8019, USA}
\date{\today }

\begin{abstract}
Quark model intrinsic wave functions of highly energetic pions in the reaction $\pppi$ are subjected to a relativistic treatment. 
The annihilation is described in a constituent quark model with $A2$ and $R2$ flavor-flux topology and the annihilated 
quark-antiquark pairs are in $\p$ and $\s$ states. 
We study the effects of pure Lorentz transformations on the antiquark and quark spatial wave functions and their
respective spinors in the pion. The modified quark geometry of the pion has considerable impact on the angular dependence 
of the annihilation mechanisms. 
\pacs{12.39.Jh, 13.75.Cs, 21.30.Fe, 25.43.+t}
\end{abstract}

\maketitle

\section{Introduction\label{intro}}

The LEAR experiments \cite{hasan} on $\pppi$ and $\ppkk$, which yielded a large and very accurate data set 
for differential cross sections $\dsig$ and analyzing powers $\ana$ from 360 to 1550 MeV/$c$, have until very 
recently \cite{elbennich1} resisted a satisfying comparison with theoretical models. In particular, the 
observable $\ana$ exhibits a characteristic double-dip structure which is hard to reproduce. Another prominent 
feature of the $\ana$ data is a shift of the asymmetry from predominantly negative values at lower momenta 
toward positive values at higher momenta. As already 
mentioned in \cite{elbennich1,kloet3}, this indicates the presence of several partial waves while most model 
calculations \cite{moussalam,mull1,mull2,kloet1,kloet2,yan9699} lead to amplitudes dominated by total angular
momentum $J=0$ and $J=1$. The reason for this feature of all models is a rather short range annihilation mechanism, 
either driven by  baryonic exchange in the $t$ channel \cite{moussalam,mull1,mull2,yan9699} or by overlap of 
quark and antiquark wave functions of the proton and antiproton as in \cite{kloet1,kloet2}.

One way to improve models would hence be to increase the annihilation range. In Ref.~\cite{elbennich1}, this problem 
is addressed within the framework of a constituent quark model. For a summary of the $\bar pp$ annihilation in a 
quark model see the review by Dover {\em et al\/}. \cite{dover}. In paper \cite{elbennich1}, the radii of 
the  proton, antiproton and pion were readjusted so they coincide with the ones obtained from measurements of the 
respective electric form factors \cite{amendolia,bo75} rather than with the considerably smaller constituent $qqq$ 
and $\bar qq$ quark core  radii. The latter radii, which ignore the hadronic $\bar qq$ cloud, were used in earlier 
attempts to reproduce the LEAR observables \cite{kloet1,kloet2}. In the model calculations, this increase in radius 
is directly related to the Gaussian description of the intrinsic quark wave functions. In the overlap integral 
over the quark coordinates, the wider Gaussian wave functions provide in turn a larger overlap of the quarks and 
antiquarks and therefore a larger range of the annihilation mechanism. In fact, the radii in Ref.~\cite{elbennich1} 
were obtained in a fit to $\dsig$ and $\ana$ data mentioned above and are about 7\% larger than the charge radii 
derived from the electric form factors. This effect, along with a fine-tuned final-state interaction of the $\pi^-\pi^+$ 
pair, improves the reproduction of the observables $\dsig$ and $\ana$, and in particular of the double-dip structure of 
the latter one, considerably. 

As was pointed out in Ref.~\cite{elbennich1}, there is another reason to believe that the geometry of the final 
pions is relevant to higher partial wave ($J>1$) contributions to the total amplitude. At the center-of-mass (c.m.) 
energies $\sqrt{s}$ considered in the LEAR experiment $\pppi$, final pions are produced with kinetic energies much 
larger than the pion rest mass. For example, for an antiproton beam with $\plab=800$ MeV/$c$ the total c.m. energy is
$E_\cm=\sqrt{s}\simeq 2020$ MeV. This translates into a relativistic factor $\gamma=E_\cm/2m_{\pi}=7.2$. 
In the quark model, the pions are described by Gaussian spheres in the rest frame of each pion. 
Yet, the transition amplitudes and implicitly the observables of the $\pppi$ reaction are calculated in the 
c.m. frame. In the c.m. frame the outgoing pions, due to their large kinetic energy, will be shaped like pancakes.
This will alter the angular dependence of the annihilation mechanism. 

This modification of the pions, due to Lorentz transformed Gaussian wave functions, was investigated previously by Maruyama 
{\em et al.\/} \cite{maruyama} in the annihilation of $\bar pp$ atoms at rest into two mesons. The authors of Ref.~\cite{maruyama}
found that the two-pion production is enhanced upon including Lorentz contraction effects. However, only branching ratios 
were discussed and the impact of these Lorentz contractions on the angular dependence of the annihilation $\pppi$ at positive
energies remained unknown.

In this paper, we will show that the pion deformation affects the overlap of intrinsic pion, proton, and antiproton wave 
functions. It was in this spirit that in Ref.~\cite{elbennich1} the width of the Gaussian wave function was increased not 
merely to reproduce the measured charge radii, but also to mock up an altered overlap of the quarks and antiquarks implied 
by relativistic considerations. In the following, we will carry out the Lorentz transformation which is a twofold task: 
the pions are $\bar qq$ pairs described by the usual spinors of the free Dirac equation times a radial Gaussian function 
to account for their confinement within the pion. Hence, the Lorentz transformation is effected on both spin and Gaussian 
components of the wave functions. Using the relevant Feynman diagrams, we analyze the relativistic effects on the 
amplitudes and compare them with previous results \cite{kloet2}.

\section{Lorentz transformed pion intrinsic wave functions}

The reaction $\pppi$ is described in the c.m. frame. Therefore, given the large kinetic energy of the final-state pions, 
we must Lorentz transform each pion intrinsic wave functions from the pion rest frame to the c.m. frame. As mentioned before, 
this will affect the Gaussian radial part of the wave function which will be distorted along the boost direction into an 
ellipsoid. We begin with an $r$-space realization of the pion wave function and first concentrate on its spatial component. 
The proton and antiproton are described by similar Gaussian functions but given their much smaller kinetic energy and larger 
mass their intrinsic distortions are ignored. The antiquark and quark coordinates in each pion transform according to
\begin{equation}
  r'=l^{-1}r .
\end{equation}
We split the quark and antiquark coordinates into components parallel and perpendicular to the boost direction $\bm{\beta}=\mathbf{v}/c$,
which is related to relativistic boost factor $\gamma=(1-\bm{\beta}^2)^{-\frac{1}{2}} = E_\cm/2m_{\pi}$.
Hence, the parallel component of the separation $\rvec_i-\rvec_j$ between an antiquark-quark pair in the pion rest frame is related to 
the one in the c.m. frame with coordinates $\rvec_i'$ and $\rvec_j'$ by
\begin{equation}
  (\rvec_i-\rvec_j)^2_{_\|} = \gamma^2 (\rvec_i'-\rvec_j')^2_{_\|} .
\label{transform}
\end{equation}
Here, we have used the equal time condition $t_i'=t_j'$ in the c.m. frame. These transformations can be applied 
straightforwardly to the spatial part of the pion wave functions. Inserting the Lorentz contraction in Eq.~(\ref{transform}), 
the pion wave function suppressing spin, isospin, and color dependence, 
\begin{equation}
 \psi_\pi(\rvec_i,\rvec_j) = N_\pi \exp \Big \{ -\frac{\beta}{2}
 \sum_{\alpha=i,j}\!(\rvec_\alpha - \mathbf{R}_{\!\pi} )^2\Big \} ,
\label{pionnoboost}
\end{equation}
becomes
\begin{eqnarray}
\psi_\pi(l^{-1}\rvec_i,l^{-1}\rvec_j) = \nonumber \hspace{5cm} \\ 
  = \tilde N_\pi \exp \Big \{ -\frac{\beta}{2}\!\sum_{\alpha=i,j}\! \left [ (\rvec_\alpha-{\bf R}_{\!\pi} )_{\!_\perp}^2\! 
  +\!\gamma^2(\rvec_{\alpha}-{\bf R}_{\!\pi} )^2_{_\|}\right ]\!\Big \} 
\label{rwaveboost}
\end{eqnarray}
where $\beta$ is in this case the size parameter and $\mathbf{R}_{\!\pi}=\frac{1}{2}(\rvec_i+\rvec_j)$ is the pion coordinate.
The new normalization factor $\tilde N_\pi=\sqrt{\gamma}\,N_\pi$ comes from the condition that Eq.~(\ref{rwaveboost}) be 
normalized to unity in the c.m. In Eq.~(\ref{rwaveboost}), the spatial distortion of the pions in the c.m. of the reaction 
$\pppi$ alluded to in the introduction is shown explicitly. In $p$-space the wave function is  
{\setlength\arraycolsep{1pt}
\begin{eqnarray}
\lefteqn{ \psi_\pi(l^{-1}\pvec_i,l^{-1}\pvec_j) =  \tilde N_\pi (4\pi/\beta)^{\!\frac{3}{2}}/\gamma\, \times } \nonumber \\ 
 & \times & \exp \Big \{\!-\frac{1}{2\beta}\!\sum_{\alpha=i,j}\!\Big [ (\pvec_\alpha-\mbox{$\frac{1}{2}$}
   {\bf P}_{\!\pi} )_{\!_\perp}^2\!+\!{\frac{1}{\gamma^2}}(\pvec_{\alpha}-\mbox{$\frac{1}{2}$}{\bf P}_{\!\pi} )^2_{_\|}
  \Big ] \Big \}  \nonumber  \\
 & = &\frac{N_\pi}{\gamma^{\!\frac{1}{2}}} \Big(\frac{4\pi}{\beta}\Big)^{\!\!\frac{3}{2}}
   \exp \Big \{ -\frac{1}{4\beta} \Big [ (\pvec_i-\pvec_j)^2 + \nonumber \\
 & & \hspace*{2cm} +\,\left (\frac{1}{\gamma^2}-1\right ) \!
 \left ( [\pvec_i-\pvec_j]\!\cdot\!{\bf \hat P}_{\!\pi}\right )^{\!2}\Big ]\Big\} .
\label{expboost}
\end{eqnarray}}

\begin{figure}[t]
\includegraphics[width=7.4cm, height=6.3cm]{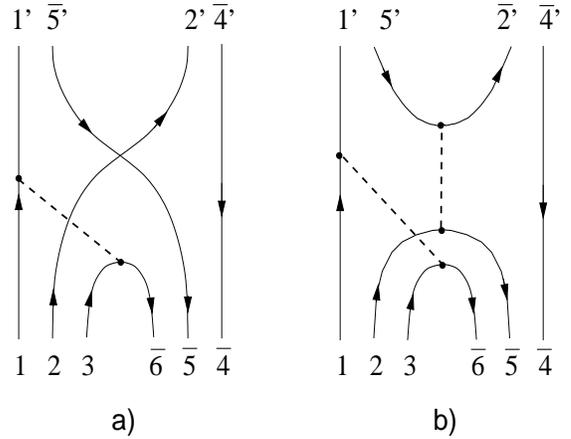}
\caption{{\bf a)} {\em Rearrangement diagram} $R2$ and {\bf b)} {\em Annihilation diagram} $A2$.
                   The numbers with bars denote antiquarks, those without bars the quarks while the dashed lines 
                   represent either the exchange of the effective ``vacuum'' $\p$ or ``gluon'' $\s$ state.}                   
\label{diagrams}
\end{figure}

\noindent
The momenta $\pvec_i$ and $\pvec_j$ are the quark and antiquark momenta, while $\mathbf{P}_\pi=\pvec_i+\pvec_j$ is the pion momentum
which is parallel to the boost direction. With these Lorentz transformed wave functions, we are now in the position to evaluate the 
$\s$ and $\p$ amplitudes taking into account the boosted pions. We will consider two types of commonly used diagrams, the 
{\em rearrangement\/} $R2$ and {\em annihilation\/} $A2$ diagrams as depicted in Fig.~\ref{diagrams}.

\section{\label{secthree}Effects of deformed pions on the annihilation amplitudes}

With the Lorentz transformed pion intrinsic wave functions, we can recalculate the $T$ matrix elements of Refs.~\cite{kloet1,kloet2} 
for the annihilation reaction $\pppi$. Here, we will do this first for the $R2$ diagrams (which we used exclusively in a previous 
paper \cite{elbennich1}). The transition operators $T(\p)$ and $T(\s)$ are obtained from integrating out the quark and antiquark 
momenta. In the following, ${\mathbf P}_p$, ${\mathbf P}_{\bar p}$, ${\mathbf P}_{\pi^-}$ and ${\mathbf P}_{\pi^+}$ are the $p$, 
$\bar p$, $\pi^-$, and $\pi^+$  momenta, respectively. For the $R2$ diagram of Fig.~\ref{diagrams}a, the transition operators for 
$\p$ as well as $\s$ are obtained from the integral: 

\begin{widetext}
{\setlength\arraycolsep{2pt}
\begin{eqnarray}
 \hat T_{R2} (\p,\s) & = & \mathcal{N} \int\!\dvec_1\,\dvec_2\,\dvec_3\,\dvec_4\,\dvec_5\,\dvec_6\,\dvec_1'\,\dvec_2'\,\dvec_4'\,\dvec_5'\,\,
  \delta (\pvec_1'+\pvec_5'-{\mathbf P}_{\pi^-})\,\delta (\pvec_2' +\pvec_4' -{\mathbf P}_{\pi^+}) \times \nonumber \\
  & \times & \delta (\pvec_1 +\pvec_2 +\pvec_3 -{\mathbf P}_p)\,\delta (\pvec_4 +\pvec_5 +\pvec_6-{\mathbf P}_{\bar p})\,
  \delta (\pvec_6 +\pvec_3 +\pvec_1 -\pvec_1')\,\delta (\pvec_2'-\pvec_2)\, \delta (\pvec_4'-\pvec_4)\,\delta (\pvec_5'-\pvec_5)
  \times  \nonumber \\
  & \times & \exp \left \{ -\frac{1}{4\beta} \left [ (\pvec_1'-\pvec_5')^2+(\pvec_2'-\pvec_4')^2+ 
  \left ( \frac{1}{\gamma^2}-1 \right ) \left \{ \left [ (\pvec_2'-\pvec_4')\!\cdot\!{\bf \hat P}_{\!\pi} \right ]^2 + 
  \left [ (\pvec_1'-\pvec_5')\!\cdot\!{\bf \hat P}_{\!\pi}\right ]^2 \right \} \right ] \right \} \times \nonumber \\
  & \times & \exp \left \{ -\frac{1}{3\alpha} \left (\sum_{i=1}^3(\pvec_i-{\mathbf P}_p)^2
    +\sum_{i=4}^6(\pvec_i-{\mathbf P}_{\bar p})^2 \right ) \right \}\, \hat V_{R2}(\p,\s) , 
\label{toperator} 
\end{eqnarray}}
\end{widetext}
where we have absorbed the relativistic factor $\gamma$, the normalization factor $N_\pi$, and the size parameter $\beta$ 
in Eq.~(\ref{expboost}) (and similarly $N_p$, $N_{\bar p}$, and the size parameter $\alpha$ in the $\bar pp$ wave function) 
into an overall normalization $\mathcal{N}$.

The two mechanisms differ only in the operators $\hat V_{R2}(\p)$ and $\hat V_{R2}(\s)$ stemming from the respective annihilation 
mechanisms:
\begin{eqnarray}
 \hat V_{R2}(\p) & = & \bm{\sigma}_{36}\!\cdot\!(\pvec_6 - \pvec_3 )\label{R2operator1} \\
 \hat V_{R2}(\s) & = & 2\,\bm{\sigma}_{36}\!\cdot\!\pvec_1+\,i\, 
  (\bm{\sigma}_{36}\!\times\!\bm{\sigma}_{1'1})\!\cdot\!(\pvec_1-\pvec_{1'}). \label{R2operator2}
\end{eqnarray}

The exponential parts in Eq.~(\ref{toperator}) originate in the $p$-space component of the boosted pion wave functions of 
Eq.~(\ref{expboost}), and the baryonic (non-boosted) wave functions, respectively. The spin-momentum dependency in Eqs.~(\ref{R2operator1})
and (\ref{R2operator2}) comes from the computation of the $R2$ diagrams using the Dirac spinors. The indices on the Pauli matrices 
denote the spinor subspace in which they act. For instance, the antiquark-quark pair $\bar q_6q_3$ is annihilated and momentum 
$\pvec_3+\pvec_6$ is transferred from this vertex to quark $q_1$. 
 
After Fourier transform into $r$-space, Eq.~(\ref{toperator}) becomes for the $\p$ mechanism   
\begin{eqnarray}
\lefteqn{\hspace{-4mm} \hat T_{R2}(\p) = i\mathcal{N} [A_V \bm{\sigma}\!\cdot\!\mathbf{R'}+B_V \bm{\sigma}\!\cdot\!\mathbf{R} 
        + C_V (\bm{\sigma}\!\cdot\!\mathbf{\hat R'})\,R\cos\theta ]} \nonumber \\
      & \times & \hspace{-1mm} \exp \{A R'^2 + B R^2 + C RR'\cos\theta + D R^2\cos^2\theta \}. 
\label{exp1}
\end{eqnarray}
The various coefficients will be given below. Since the $\s$ mechanism in Eq.~(\ref{R2operator2}) contains two terms, we split 
their contributions conveniently into a {\em longitudinal\/} and a {\em transversal\/} part. It will be seen that they give
rise to different selection rules. The transition operator is thus   
{\setlength\arraycolsep{2pt}
\begin{eqnarray}
\hat T_{R2}(\s) & = &  \mathcal{N} [ A_L i\bm{\sigma}\!\cdot\!\mathbf{R'} + B_L i\bm{\sigma}\!\cdot\!\mathbf{R} 
       + C_L i\bm{\sigma}\!\cdot\!\mathbf{\hat R'}\,R\cos\theta    \nonumber \\
      & + & A_T\, \bm{\sigma}\!\cdot\!\mathbf{R'} + B_T \bm{\sigma}\!\cdot\!\mathbf{R} 
          + C_T (\bm{\sigma}\!\cdot\!\mathbf{\hat R'})\,R\cos\theta ] \nonumber \\
       & &\hspace{-14mm}\times \exp \{A R'^2 + B R^2 + C RR'\!\cos\theta + D R^2\!\cos^2\theta \}. 
\label{exp2}
\end{eqnarray}}

\noindent
The relative $\pi^-\pi^+$ and antiproton-proton coordinates are $\mathbf{R'}=\mathbf{R}_{\pi^-}\!\!-\mathbf{R}_{\pi^+}$ 
and $\mathbf{R}=\mathbf{R}_{\bar p}-\mathbf{R}_p$, respectively. The angle $\theta$ is between the relative $\pi^-\pi^+$
coordinate $\mathbf{R'}$ and antiproton-proton coordinate $\mathbf{R}$ in the c.m. system. If compared with the results of 
Ref.~\cite{kloet2}, one observes the appearance of additional terms in both transition operators, namely the $D$ term in the 
exponent and the $C_V$, $C_L$, and $C_T$ terms. These new terms introduce manifest angular dependence. 
More precisely, the relativistic pion wave functions lead to additional angular dependence in the form of quadratic cosine 
terms in the exponentials and linear cosine terms in the coefficients of the spin operators. Note that aside of the explicit 
non-locality, the transition operators $\hat T_{R2}(\p)$ and $\hat T_{R2}(\s)$ are now energy dependent via the Lorentz boost 
factor $\gamma$. This $\gamma$-dependence is non-trivial and affects the angular dependence of the transition operators in the sense that 
the various terms mentioned above behave differently for increasing values of $\gamma$. Also, the magnitude of these terms differs
immensely depending on whether $\gamma$ is small ($\gamma \simeq 1-2$) or large ($\gamma \simeq 5-8$). For large 
$\gamma$ the $D$ term dominates the exponential part and, for example, $C_V\gg A_V, B_V$ and $C_L\gg A_L, B_L$,
however,  $C_T , A_T ,$ and $B_T$ are of the same order of magnitude.

The above results for the transition operators are valid for the diagram in Fig.~\ref{diagrams}a. Summing over all 
possible permutations of Fig.~\ref{diagrams}a, where for example $\pvec_3+\pvec_6$ is transferred to quark $q_2$ or one of the 
antiquarks $\bar q_4,\bar q_5$, yields the same symmetry properties as in Ref.~\cite{kloet2}, as expected. In particular, the {\em vacuum\/} 
coefficients $A_V$, $B_V$, and $C_V$ add up such that $\hat T_{R2}(\p)$ contributes only to $l_{\pi\pi}=0,2,4,...$ and 
$l_{\bar pp}=1,3,5,...$ and hence acts in $\bar pp$ states with $J^\pi=0^+,2^+,4^+,...$ waves. The longitudinal part of the 
transition amplitude $\hat T_{R2}(\s)$, with the coefficients $A_L$, $B_L$, and $C_L$, bears the same symmetry properties and 
therefore acts in the same waves as the ``vacuum'' mechanism. The transversal component with $A_T$, $B_T$, and $C_T$, on the 
other hand, contributes to $l_{\pi\pi}=1,3,5,...$  and $l_{\bar pp}=0,2,4,...$ and acts therefore in $\bar pp$ states with 
$J^\pi=1^-,3^-,5^-,...$ waves.

The complete form of the $R2$ transition operators for the vacuum $\p$ amplitude when summing over all permutations is

\begin{eqnarray}
\lefteqn{\hat T^{\mbox{\tiny tot.}}_{R2}(\p) =} \nonumber \\
& =& i\mathcal{N}\Big [ A_V \bm{\sigma}\!\cdot\!\mathbf{R'}\sinh(C\,\mathbf{R}\!\cdot\!\mathbf{R'}) +
 B_V \bm{\sigma}\!\cdot\!\mathbf{R}\cosh(C\,\mathbf{R}\!\cdot\!\mathbf{R'})\,  \nonumber \\
& & + \,\, C_V (\bm{\sigma}\!\cdot\!\mathbf{\hat R'})\,R\cos\theta 
 \cosh(C\,\mathbf{R}\!\cdot\!\mathbf{R'})\Big ]\times \nonumber \\
& &\times \, \exp \{A\mathbf{R'}^2+B\mathbf{R}^2+D \mathbf{R}^2\cos^2\theta \} .
\label{tot1} 
\end{eqnarray}
The total $\s$ amplitude for the longitudinal component is given by
\begin{eqnarray}
\lefteqn{\hat T^{\mbox{\tiny tot.}}_{R2}(\s^L) =} \nonumber \\
& = & i\mathcal{N}\Big [ A_L \bm{\sigma}\!\cdot\!\mathbf{R'}\sinh(C\,\mathbf{R}\!\cdot\!\mathbf{R'}) +
 B_L \bm{\sigma}\!\cdot\!\mathbf{R}\cosh(C\,\mathbf{R}\!\cdot\!\mathbf{R'})\,  \nonumber \\
& & + \,\, C_L (\bm{\sigma}\!\cdot\!\mathbf{\hat R'})\,R\cos\theta 
 \cosh(C\,\mathbf{R}\!\cdot\!\mathbf{R'})\Big ]\times \nonumber \\
& &\times \, \exp \{A\mathbf{R'}^2+B\mathbf{R}^2+D \mathbf{R}^2\cos^2\theta \} 
\label{tot2}
\end{eqnarray}
and
\begin{eqnarray}
\lefteqn{\hat T^{\mbox{\tiny tot.}}_{R2}(\s^T) =} \nonumber \\
& = & \mathcal{N}\Big [ A_T \bm{\sigma}\!\cdot\!\mathbf{R'}\cosh(C\,\mathbf{R}\!\cdot\!\mathbf{R'}) + 
 B_T \bm{\sigma}\!\cdot\!\mathbf{R}\sinh(C\,\mathbf{R}\!\cdot\!\mathbf{R'})\,  \nonumber \\
& & + \,\, C_T (\bm{\sigma}\!\cdot\!\mathbf{\hat R'})\,R\cos\theta 
 \sinh(C\,\mathbf{R}\!\cdot\!\mathbf{R'})\Big ]\times \nonumber \\
& &\times \, \exp \{A\mathbf{R'}^2+B\mathbf{R}^2+D \mathbf{R}^2\cos^2\theta \} 
\label{tot3}
\end{eqnarray}
for the transversal component.

The explicit expression for the coefficients of Eqs.~(\ref{tot1}), (\ref{tot2}), and (\ref{tot3}) is:
\begin{subequations}
\begin{eqnarray}
 A & = & -\frac{\alpha(5\alpha+4\beta\gamma^2)}{2(4\alpha+3\beta\gamma^2)} \\
 B & = & -\frac{3(7\alpha^2 +18\alpha\beta\gamma^2+9\beta^2\gamma^4)}{8(4\alpha+3\beta\gamma^2)}-D \\
 C & = & -\frac{3\alpha(\alpha+\beta\gamma^2)}{2(4\alpha+3\beta\gamma^2)} \\ 
 D & = & -\frac{9\beta(\gamma^2-1)}{8}\!\left \{ 1+\frac{\alpha^2}{(5\alpha+4\beta)(5\alpha+4\beta\gamma^2)}\right\}
\hspace{8mm}
\end{eqnarray}
\end{subequations}
\begin{subequations}
\begin{eqnarray}
 A_V & = & \frac{\alpha(\alpha+\beta\gamma^2)}{4\alpha+3\beta\gamma^2}  \\
 B_V & = & \frac{3(\alpha+\beta\gamma^2)(5\alpha+3\beta\gamma^2)}{2(4\alpha+3\beta\gamma^2)} - C_V  \\
 C_V & = & \frac{3\beta(\gamma^2-1)}{2}\left \{ 1+\frac{\alpha^2}{(5\alpha+4\beta)(5\alpha+4\beta\gamma^2)} \right \} 
\hspace{8mm}
\end{eqnarray}
\end{subequations}
\begin{subequations}
\begin{eqnarray}
 A_L & = & -A_V  \\
 B_L & = & \frac{9(\alpha+\beta\gamma^2)^2}{2(4\alpha+3\beta\gamma^2)}-C_L \\
 C_L & = & \frac{3\beta(\gamma^2-1)}{2}\left \{ 1-\frac{\alpha^2}{(5\alpha+4\beta)(5\alpha+4\beta\gamma^2)} \right \}
\hspace{8mm}
\end{eqnarray}  
\end{subequations}
\begin{subequations}
\begin{eqnarray}
 A_T & = & -2A_V \\
 B_T & = &  \frac{3\alpha(\alpha+\beta\gamma^2)}{4\alpha+3\beta\gamma^2}-C_T \\
 C_T & = & - \frac{3\beta\alpha^2(\gamma^2-1)}{(5\alpha+4\beta)(5\alpha+4\beta\gamma^2)} .
\hspace{3.4cm}
\end{eqnarray} 
\end{subequations}

In the non-relativistic limit $\gamma\rightarrow 1$, the results of Ref.~\cite{kloet2} are recovered as 
expected. In particular, the coefficients $D$, $C_V$, $C_L$, and $C_T$ vanish in this limit.

Secondly, we consider the $A2$ diagrams, where the quarks and antiquarks are integrated out similarly to Eq.~(\ref{toperator}) 
but where the different flavor-flux topology implies a different momentum transfer. Effectively, this results into one delta 
function less than for the $R2$ diagram. This will affect the angular dependence and selection rules as will be seen shortly. 

We take into account two types of $A2$ diagrams --- one in which both $\bar q_6 q_3$ and  $\bar q_5 q_2$ pairs in 
Fig.~\ref{diagrams}b annihilate into a $\s$ state, while in the other diagram the $\bar q_6 q_3$ pair annihilates into the 
``vacuum'' $\p$ state and the $\bar q_5 q_2$ pair annihilates into the ``gluon'' $\s$ state followed by the creation of a 
$\bar q_{5'} q_{2'}$ pair. We do not take into account the double $\p$ annihilation since the resulting operator is of the 
order $\mathcal{O}[(\pvec/m)^3]$. The integrals for the two cases are
\begin{widetext}
{\setlength\arraycolsep{2pt}
\begin{eqnarray}
\hat T_{A2}(\mbox{$\frac{\p}{\s}$},\mbox{$\frac{\s}{\s}$}) & = & 
  \mathcal{N} \int\!\dvec_1\,\dvec_2\,\dvec_3\,\dvec_4\,\dvec_5\,\dvec_6\,\dvec_1'\,\dvec_2'\,\dvec_4'\,\dvec_5'\,\,
  \delta (\pvec_1'+\pvec_5'-{\mathbf P}_{\pi^-})\,\delta (\pvec_2' +\pvec_4' -{\mathbf P}_{\pi^+}) \times \nonumber \\
  & \times & \delta (\pvec_1 +\pvec_2 +\pvec_3 -{\mathbf P}_p)\,\delta (\pvec_4 +\pvec_5 +\pvec_6-{\mathbf P}_{\bar p})\,
  \delta (\pvec_6 +\pvec_3 +\pvec_1 -\pvec_1')\,\delta (\pvec_2'+\pvec_5'-\pvec_2-\pvec_5)\,\delta (\pvec_4'-\pvec_4)
  \times  \nonumber \\
  & \times & \exp \left \{ -\frac{1}{4\beta} \left [ (\pvec_1'-\pvec_5')^2+(\pvec_2'-\pvec_4')^2+ 
  \left ( \frac{1}{\gamma^2}-1 \right ) \left \{ \left [ (\pvec_2'-\pvec_4')\!\cdot\!{\bf \hat P}_{\!\pi} \right ]^2 + 
  \left [ (\pvec_1'-\pvec_5')\!\cdot\!{\bf \hat P}_{\!\pi}\right ]^2 \right \} \right ] \right \} \times \nonumber \\
  & \times & \exp \left \{ -\frac{1}{3\alpha} \left (\sum_{i=1}^3(\pvec_i-{\mathbf P}_p)^2
    +\sum_{i=4}^6(\pvec_i-{\mathbf P}_{\bar p})^2 \right ) \right \}\, \hat V_{A2}(\mbox{$\frac{\p}{\s}$},\mbox{$\frac{\s}{\s}$}) .
\label{toperatorA2} 
\end{eqnarray}}
\end{widetext} 
The two cases differ only in the operators $\hat V_{A2}$ according to the type of annihilation mechanisms used:
\begin{equation}
\hat V_{A2}(\mbox{$\frac{\p}{\s}$}) =   \hat V_{R2}(\p) \times \bm{\sigma}_{25}\!\cdot\!\bm{\sigma}_{2'5'} 
\label{A2operator1} 
\end{equation}
\begin{equation}
 \hat V_{A2}(\mbox{$\frac{\s}{\s}$})  =  \hat V_{R2}(\s) \times \bm{\sigma}_{25}\!\cdot\!\bm{\sigma}_{2'5'} .
\label{A2operator2}
\end{equation}
The notation is the same as for the $R2$ diagrams. Following Fig.~\ref{diagrams}b, momentum is transferred from 
the annihilated $\bar q_6 q_3$ pair to quark $q_1$ (but can also be transferred to $\bar q_4$) and from the 
$\bar q_5 q_2$ pair to the created pair $\bar q_{5'} q_{2'}$. In $r$-space the $A2$ transition operators for the
mixed $\p$ and $\s$ vertices are
\begin{eqnarray}
\hat T_{A2}(\mbox{$\frac{\p}{\s}$}) & = &
 \mathcal{N}[\bm{\sigma}\!\cdot\!\bm{\nabla}_{\!R'} + (A_V + B_V)\, i \bm{\sigma}\!\cdot\!\mathbf{R} ] \times \nonumber \\ 
 & \times &  \delta (3\mathbf{R}/2-\mathbf{R}')\, \exp \{ (B+D)R^2 \}  \hspace*{1.5cm}
\label{a2amp1} 
\end{eqnarray}
and for two $\s$ annihilation vertices
\begin{eqnarray}
\hat T_{A2}(\mbox{$\frac{\s}{\s}$}) & = & \mathcal{N}[\bm{\sigma}\!\cdot\!\bm{\nabla}_{\!R'}+
 (A_L + B_L)\,i\bm{\sigma}\!\cdot\!\mathbf{R} + \nonumber \\ 
 & + & (A_T + B_T)\bm{\sigma}\!\cdot\!\mathbf{R}] \times \nonumber \\ 
 &\times & \delta (3\mathbf{R}/2-\mathbf{R}')\,\exp \{ (B+D)R^2 \}. 
\label{a2amp2}
\end{eqnarray}
For convenience the relativistic terms $B_V$, $B_L$, $B_T$ in the spin coefficients and $D$ in the exponent are written separately 
from the corresponding non-relativistic terms $A_V$, $A_L$, $A_T$, and $B$. All are listed explicitly below. 

It is a striking feature of the $A2$ amplitudes that they depend solely on the relative proton-antiproton $\mathbf{R}$ 
vector but {\em not} on the relative orientation of $\mathbf{R}$ and $\mathbf{R}'$. This result is an implicit consequence
of Fourier transforming the $A2$ amplitudes to $r$-space which yields a delta function $\delta (3\mathbf{R}/2-\mathbf{R}')$. 
One does obtain additional terms due to the relativistic corrections in the pion wave functions, yet these are simpler than for 
the $R2$ diagrams and they do not introduce extra angular dependence. Again certain terms, namely $D$, $B_V$, 
$B_L$, and $B_T$, are energy dependent via the boost factor $\gamma$. In particular, the exponential factor $D$ is very large 
(for $\gamma\simeq 7$ it is about fifteen times larger than $B$) and acts like a short-range angle-independent cut-off.

Since the $A2$ diagrams were not discussed in previous work \cite{kloet2}, we take the opportunity to analyze their symmetry 
properties. Taking into account spin-flavor and color matrix elements, the amplitudes in Eqs.~(\ref{a2amp1}) and (\ref{a2amp2}) 
are identical for the permutation of the $A2$ diagram (Fig.~\ref{diagrams}b) in which a $\p$ or $\s$ state resulting
from the $\bar q_6q_3$ annihilation is exchanged with $\bar q_4$ rather than with $q_1$. This means that the complete form of
$\hat T_{A2}^{\mbox{\tiny tot.}}$ remains as in Eqs.~(\ref{a2amp1}) and (\ref{a2amp2}) and that
$\hat T_{A2}(\mbox{$\frac{\p}{\s}$})$ and $\hat T_{A2}(\mbox{$\frac{\s}{\s}$})$ are {\em odd\/} in $\mathbf{R}$ and $\mathbf{R}'$. 
Because of the delta function $\delta (3\mathbf{R}/2-\mathbf{R}')$ all $A2$ operators, regardless of the annihilation mechanisms 
employed, act in ($l_{\pi\pi}\neq l_{\bar pp}$) $J^\pi=0^+,1^-,2^+,3^-,4^+,5^-,...$ waves. 

We end this section with the expressions for the coefficients of the transition amplitudes (\ref{a2amp1})
and (\ref{a2amp2}):
{\setlength\arraycolsep{1pt}
\begin{subequations}
\begin{eqnarray}
 B &=& -\frac{3(28\alpha^2+36\alpha\beta+9\beta^2)}{8(4\alpha+3\beta)} \\
 D &=& - 9\beta(\gamma^2-1)\left \{ 1+\frac{4\alpha^2}{(4\alpha+3\beta)(4\alpha+3\beta\gamma^2)} \right \}
\hspace{9mm}
\end{eqnarray}
\end{subequations}
\begin{subequations}
\begin{eqnarray}
 A_V &=& \frac{3(12\alpha^2+14\alpha\beta+3\beta^2)}{2(4\alpha+3\beta)} \\
 B_V &=& \frac{3\beta(\gamma^2-1)}{2}\left \{ 1+\frac{4\alpha^2}{(4\alpha+3\beta)(4\alpha+3\beta\gamma^2)} \right \}
\hspace{9mm}
\end{eqnarray}
\end{subequations}
\begin{subequations}
\begin{eqnarray}
 A_L &= &  \frac{3(2\alpha+\beta)}{2} \\
 B_L &=& \frac{3\beta(\gamma^2-1)}{2}
\hspace{5.4cm}
\end{eqnarray}
\end{subequations}
\begin{subequations}
\begin{eqnarray}
A_T & = & -\frac{6\alpha(\alpha+\beta)}{4\alpha+3\beta}   \\
B_T & = & -\frac{6\alpha^2\beta(\gamma^2-1)}{(4\alpha+3\beta)(4\alpha+3\beta\gamma^2)}~.
\hspace{3.2cm}
\end{eqnarray}
\end{subequations}}\noindent
Clearly, $D$, $B_V$, $B_L$, and $B_T$ vanish in the non-relativistic limit $\gamma \rightarrow 1$.

\section{Geometric interpretation}

\begin{figure}[b]
 \includegraphics[width=0.45\textwidth, height=0.2\textwidth]{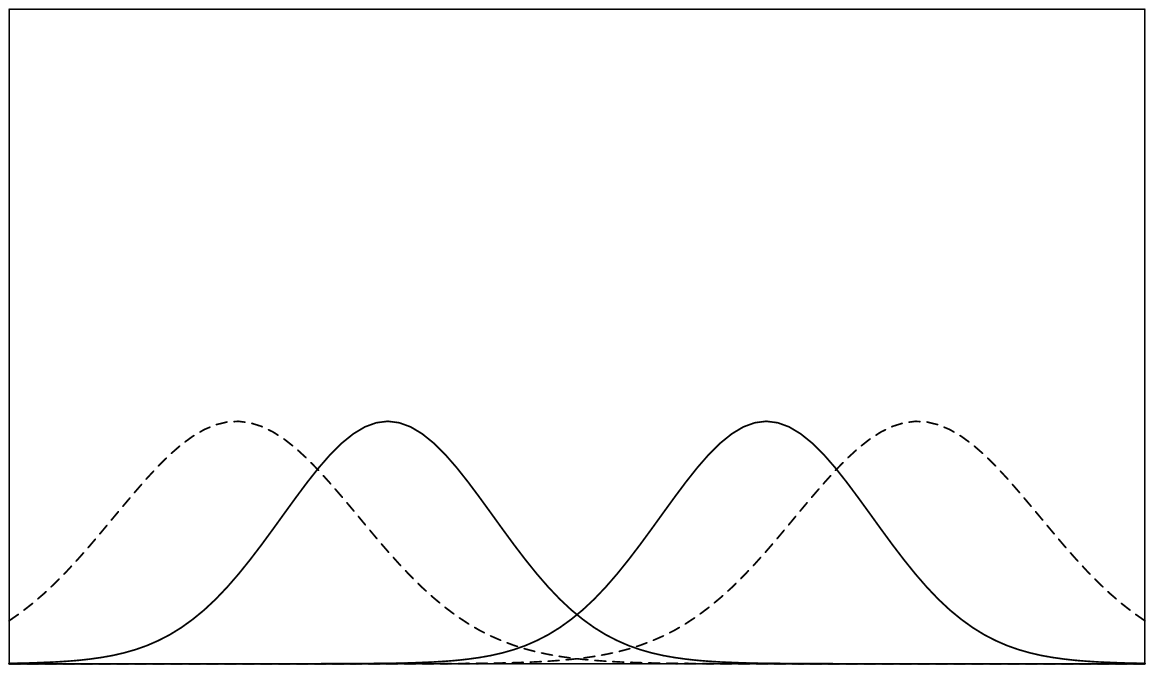}
 \includegraphics[width=0.45\textwidth, height=0.2\textwidth]{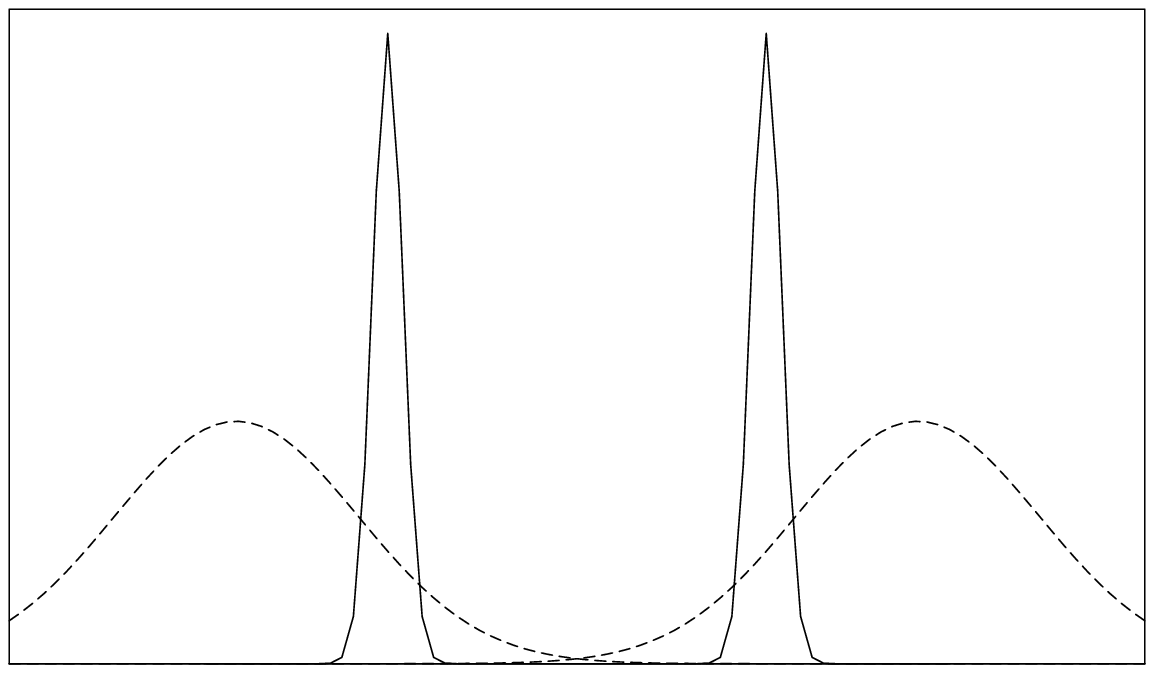}
\caption{The upper graph shows a two-dimensional projection of the intrinsic pion wave functions (solid lines) given by
 Eq.~(\ref{pionnoboost}) and (anti)proton wave functions (dashed lines). The distance between the peaks of the solid-line Gaussians 
represents the relative pion coordinate $\mathbf{R}'=\mathbf{R}_{\pi^-}\!\!-\mathbf{R}_{\pi^+}$ whereas the distance between the peaks 
of the dashed-line Gaussians is the relative $\bar pp$ coordinate $\mathbf{R}=\mathbf{R}_{\bar p}-\mathbf{R}_p$. The distances are the 
same in both graphs, however, in the lower graph the boosted intrinsic pion wave function is given by Eq.~(\ref{rwaveboost}) with $\gamma=7$.}
\label{twogauss}
\end{figure}

It is important to understand the geometric implications of deformed pion wave functions as required by relativity and their
relation to the $\bar pp$ annihilation range. In Sec.~\ref{intro} we recall that in a previous (non-relativistic) attempt to 
explain the LEAR data on $\pppi$ \cite{elbennich1}, an increase of proton, antiproton, and pion radii effectively augments 
the range of the transition operators $\hat T_{R2}$ and $\hat T_{A2}$. It is therefore surprising that one can obtain any improvement from 
the Lorentz contraction of pions \cite{elbennich2}. After all,  we ``shrink'' the intrinsic pion wave function in the boost direction, which 
amounts to less overlap of the quark and antiquark wave functions. The graphic description in Fig.~\ref{twogauss} visualizes this. The two 
intrinsic pion wave functions are plotted at a fixed distance $|\mathbf{R'}|$ from each other and the centers of the (also Gaussian) 
proton and antiproton intrinsic wave functions are separated by a distance $|\mathbf{R}|$. One sees that when the pion wave functions are 
Lorentz contracted in the c.m., the overlap of pions, proton, and antiproton (and of a transition operator depending on the annihilation 
mechanism, which we omit here for simplicity) decreases considerably and so does their contribution to the annihilation amplitude. Since 
the total probability is conserved, the Gaussian wave functions of the pions are peaked higher in the lower graph. 

\begin{figure}[t]
 \includegraphics[scale=0.55]{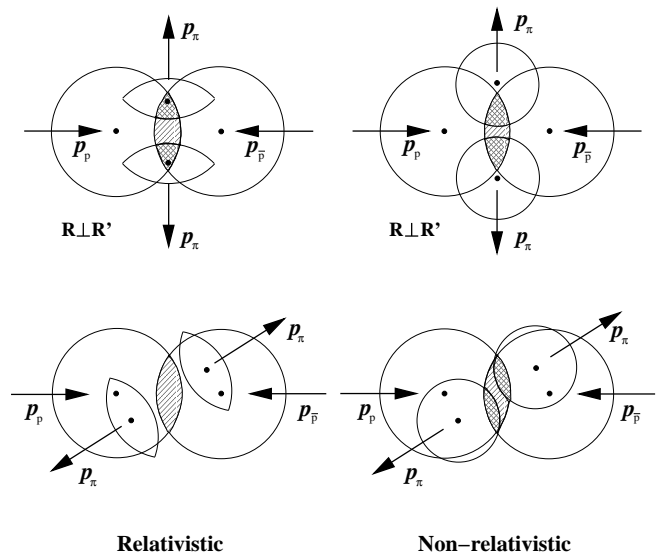}
\caption{Sketch of the overlap in the reaction $\pppi$ in the c.m. system at two different angles $\theta_{\mbox{\tiny c.m.}}$ 
between the incoming $\bar pp$ and the outgoing $\pi\pi$ pairs. In the left column, the pions are Lorentz contracted. 
while in the right column they are described by spherical wave functions. One sees that the angular dependence
of the overlaps at same distances $\mathbf{R}'$ and $\mathbf{R}$ strongly differs in relativistic and non-relativistic case.}
\label{pp2pi}
\end{figure}

As discussed in Sec.~\ref{secthree}, the $R2$ transition operators acquire new terms which introduce additional angular dependence. 
The effect of the $D$ term in Eqs.~(\ref{exp1}) and (\ref{exp2}), for example, is strongly angle dependent --- for $\mathbf{R}$ parallel 
or anti-parallel to $\mathbf{R'}$, the $\mathbf{R}$-range of $\hat T_{R2}$, is drastically reduced due to large $D$ values, whereas for 
$\mathbf{R}\perp\mathbf{R}'$ this cut-off effect is absent. This angular dependence is schematically depicted in Fig.~\ref{pp2pi}, 
where $\mathbf{R'}$ and $\mathbf{R}$ have fixed magnitudes. For $\mathbf{R}\perp\mathbf{R}'$ in the top panels one sees that even 
though the pions are Lorentz contracted, each pion still overlaps with the $\bar pp$ pair as in the non-relativistic case. This 
overlap vanishes at small angles $\theta$ (see lower panels in Fig.~\ref{pp2pi}) and at $\theta=0~(\theta=\pi)$ when $\mathbf{R'}$ is 
(anti)parallel to $\mathbf{R}$. This illustrates that the angle dependence of the overlap in the relativistic and non-relativistic case are 
quite different. That the least overlap occurs for $\mathbf{R}\,\|\,\mathbf{R}'$ is expressed in the transition operators $\hat T_{R2}$ by 
the $D R^2 \cos^2\theta$ term in the exponent of Eqs.~(\ref{tot1}), (\ref{tot2}), and (\ref{tot3}). The overlap is minimal at these angles 
as are the resulting amplitudes $\hat T_{R2}$. 

The angular momentum content of the scattering matrix elements $T^J$ is obtained by sandwiching the transition operators between the 
initial-state $\bar pp$ and the final-state $\pi^-\pi^+$ wave functions, 
\begin{equation}
\hspace*{-1mm} T^J =\int\! \mathrm{d}\mathbf{R} \mathrm{d}\mathbf{R'} \phi_{\pi\pi}^J(\mathbf{R'})\,
 \hat T_{R2/A2}(\mathbf{R},\mathbf{R'})\,\Psi_{\bar pp}^{J=l\pm 1}(\mathbf{R}) ,
\label{int}
\end{equation}
where $\phi_{\pi\pi}^J(\mathbf{R'})$ and $\Psi_{\bar pp}^{J=l\pm1}(\mathbf{R})$ still contain the usual angular dependence
in the form of appropriate spherical harmonics. Therefore, additional angle-dependent terms in $\hat T_{R2/A2}(\mathbf{R},\mathbf{R'})$ 
enhance the contribution of higher angular momenta $J$. The resulting increase of contributions to higher partial waves $J\geq 1$ is 
the main reason for a better description of the experimental data. We will give numerical evidence for this in another communication 
\cite{elbennich2}. A partial wave analysis of the integral in Eq.~(\ref{int}) will show that the richer angular dependence 
of the $R2$ transition operators considerably increases contributions to partial waves $J=2$ and higher.

\section{Lorentz transformation of (anti)quark spinors \label{sec5}}

So far, we have solely treated the Gaussian part of the pion intrinsic wave functions. An $s$-wave pion wave function 
in momentum space, omitting flavor and color components, is given by
\begin{eqnarray}
 \psi_{\pi}(\pvec_i,\pvec_j) = N_\pi (4\pi/\beta)^{\frac{3}{2}}\, \delta (\pvec_i + \pvec_j - {\bf P}_{\!\pi})
   \nonumber\hspace{1.3cm} \\ 
   \times\, u(\pvec_i)v(\pvec_j) \exp\!\left [ -\frac{1}{2\beta} \sum_{\alpha=i,j}(\pvec_{\alpha}-\mbox{$\frac{1}{2}$}
   {\bf P}_{\!\pi} )^2\right ]\!.
\label{pion}
\end{eqnarray}
The momenta $\pvec_i$ and $\pvec_j$ denote the antiquark and quark respectively, ${\bf P}_{\!\pi}$ is the pion momentum, 
$\beta$ is the parameter that determines the size of the pion and $u(\pvec_i)$ and $v(\pvec_j)$ are the usual quark and 
antiquark Dirac spinors for particles with spin $s=\frac{1}{2}$,
\begin{subequations}
\begin{eqnarray}
 u(\pvec,m_s)& = &  \left ( \begin{array}{c} \sqrt{E+m} \\ \frac{\bm{\sigma }\cdot\pvec}{\sqrt{E+m}} 
                                   \end{array} \right ) \chi_{_{m_s}}  \\
 v(\pvec,m_s)& = &  \left ( \begin{array}{c} \frac{\bm{\sigma}\cdot\pvec}{\sqrt{E+m}} \\ \sqrt{E+m}
                                   \end{array} \right ) \chi_{_{-m_s}} (-)^{\frac{1}{2}-m_s} 
\end{eqnarray}
\label{spinors}
\end{subequations}
or in terms of helicity spinors
\begin{subequations}
\begin{eqnarray}
 u(\pvec,\lambda)& = &  \left ( \begin{array}{c} \sqrt{E+m} \\ \frac{2p\lambda}{\sqrt{E+m}} 
                                   \end{array} \right ) \chi_{_{\lambda}}(\hat\pvec)  \\
 v(\pvec,\lambda)& = &  \left ( \begin{array}{c} \frac{-2p\lambda}{\sqrt{E+m}} \\ \sqrt{E+m}
                                   \end{array} \right ) \chi_{_{-\lambda}}(\hat\pvec)  (-)^{\frac{1}{2}-\lambda} .
\end{eqnarray}
\label{helispinors}
\end{subequations}

Our approach differs from that of Maruyama {\em et al\/}. \cite{maruyama}, who use an approximation of the $1s$ quark 
wave function in the MIT bag in $r$-space as introduced in Ref.~\cite{wong,duck}. In Ref.~\cite{maruyama} the quark wavefunction is 
\begin{eqnarray}
 \psi_q({\bf r}_i) & = & [R_0^3\pi^{\frac{3}{2}}(1+3\zeta^2/2)]^{-\frac{1}{2}} \times \nonumber\\
 & \times& \exp \{-r_i^2/2R_0^2 \} \left (\!\begin{array}{c} 1 \\ i\zeta\frac{\bm{\sigma}\cdot{\bf r}_i}{R_0} \end{array}\!\right )\!\chi ,
\label{quarkwong}
\end{eqnarray}
where $R_0$ is the size parameter, which can be related to the bag radius. The parameter $\zeta$ determines the probability that the 
quark is in the {\em lower} component of the Dirac spinor and is referred to as the small relativistic component of the wave function. 
The value of $\zeta$ in the bag model has been estimated to be 0.36 for zero quark mass. The authors of Ref.~\cite{maruyama} also Lorentz 
transformed the Gaussian part of the wave function in Eq.~(\ref{quarkwong}) into the c.m. but did not apply this transformation to the 
spinor. In their analysis, Maruyama {\em et al\/}. found that the small relativistic components have little influence on the predictions 
of relative ratios of $\bar pp$ annihilation amplitudes into mesons. The effect of including the Lorentz contraction, on the other hand, 
is to enhance the production of two pions.

In Sec.~\ref{secthree}, the momentum space transition operators in Eqs.~(\ref{exp1}), (\ref{exp2}), (\ref{a2amp1}), and (\ref{a2amp2}) 
were calculated making use of just the Gaussian part of the wave function of Eq.~(\ref{pion}). In the present section 
we investigate how Lorentz boosts of quark spinors in the pion wave function of Eq.~(\ref{pion}) affect these calculations. 

In the pion rest frame $\spi$ the quark four-momentum is $p$, while in the c.m. frame $S_\cm$ of the reaction $\pppi$ the quark 
four-momentum is $p'$. They are related by the inverse Lorentz transformation $l^{-1} \equiv l^{-1}(\bm{\beta})$ as follows
\begin{equation}
  p' = l^{-1} p .
\label{lorentz}
\end{equation}
Here, $\bm{\beta} = \mathbf{v}/c$ is a vector along the pion momentum ${\bf P}_{\!\pi}$. 

Helicity states in frame $\spi$ are denoted by $|\pvec,\lambda\rangle_{S_\pi}$  while in the c.m. frame $S_\cm$ they are 
$|\pvec',\lambda'\rangle_{S_\cm}$. The two sets of states are related by 
\begin{equation}
 |\pvec',\lambda'\rangle_{S_\cm}= U(l^{-1}) |\pvec,\lambda\rangle_{S_\pi},
\label{boost}
\end{equation}
where $U(l^{-1})$ is a unitary operator effecting the inverse Lorentz transformation.
It can be shown for any Lorentz transformation $l$ that the operation of  $U(l)$ on the state $|\pvec,\lambda\rangle$ is equivalent to a 
rotation $r(l,\pvec)$ in spin space. 
For a detailed derivation of Lorentz transformations of helicity states see for instance  Ref.~\cite{leader}. The rotations $r(l,\pvec)$
that yield the c.m. states $|\pvec,\lambda \rangle_{S_\cm}$ are the Wick helicity rotations. With the appropriate $2s+1$ dimensional 
matrix representation $\D [r(l,\pvec)]$, where $s=\frac{1}{2}$ is the spin of the (anti)quark, one can rewrite Eq.~(\ref{boost}) as 
\begin{equation}
|\pvec,\lambda \rangle_{S_\pi} \stackrel{l^{-1}}{\longrightarrow} |\pvec',\lambda'\rangle_{S_\cm} = 
                               \mathcal{D}^s_{\mu\lambda'}[r(l,\pvec)] |\,l^{-1}\pvec,\mu\rangle_{S_\pi}. 
\label{pstate}
\end{equation}
A sum over $\mu$ is implied. The quark wave functions are obtained from
\begin{equation}   
  \langle 0|\psi(x)|\pvec,\lambda\rangle = \frac{1}{\sqrt{2\pi}}\,u(\pvec,\lambda)e^{-ipx} ,
\end{equation}
where $\psi(x)$ is the quark Dirac field. The correspondence $|\pvec,\lambda \rangle \longleftrightarrow u(\pvec,\lambda)$ 
between the physical state and its wave function realization yields the desired Lorentz transformation of the Dirac spinor
\begin{subequations}
\begin{equation}
  u(\pvec',\lambda' )_{S_\cm} = \D(r) u(l^{-1}\pvec,\lambda)_{S_\pi} \hspace{4mm}
\end{equation}
and similar arguments lead to
\begin{equation}
  v(\pvec',\lambda' )_{S_\cm} = \D(r^{-1}) v(l^{-1}\pvec,\lambda)_{S_\pi} .
\end{equation}
\end{subequations}
For the general case of a boost of the axes $l(\bm{\beta})$, with  $\bm{\beta} = (\beta,\theta_{\beta},\phi_{\beta})$, 
the matrices $\D$ are
\begin{equation}  
 \D = e^{i\eta(\lambda'-\lambda)}d^s_{\lambda\lambda'}(\wick) ,
\label{wickrot}
\end{equation}
where $\eta$ is a function of the angles $\theta_\beta$ and $\phi_\beta$ and the Wick angle $\wick$ \cite{leader} is given by
\begin{equation}  
  \cos \wick = \frac{\gamma}{p'}(p+\beta E\cos\delta).
\end{equation} 
Here, $E=\sqrt{\pvec^2+m^2}$ is the energy of one (anti)quark as in Eqs.~(\ref{spinors}a) and (\ref{spinors}b), 
$\gamma=E_{\cm}/2m_{\pi}$, $p=|\pvec|$, $p'=|l^{-1}\pvec|$, and \mbox{$0\leq \delta\leq \pi$} is the angle between 
$\pvec$ and $\bm{\beta}$, defined by
\begin{equation}
 \cos \delta = \frac{\bm{\beta}\cdot\pvec}{\beta p} = \frac{{\bf P}_{\!\pi}\!\cdot\pvec}{P_{\!\pi}\,p} .
\label{delta}
\end{equation}

For Lorentz boosts $l_z(\bm{\beta})$, for which the $z$-axis is along ${\bf P}_{\!\pi}$ in the c.m. system and 
$\theta_{\beta}=\phi_{\beta}=\eta=0$, the expression in Eq.~(\ref{wickrot}) simplifies to 
\begin{equation}  
  \D = d_{\lambda\lambda'}^s (\wick) .
\end{equation} 
Since $s=\frac{1}{2}$, only two matrix elements are required \cite{brink}:
\begin{subequations}
\begin{eqnarray}
   d_{1/2,1/2}^{1/2}(\wick) &=& \cos \frac{\wick}{2}\, =\, \sqrt{\frac{1}{2}+\frac{1}{2}\cos\wick} = \nonumber  \\
                            & = &   \sqrt{\frac{1}{2}+\frac{\gamma}{2p'}(p+\beta E\cos\delta)} , \label{d1} \\
 d_{-1/2,1/2}^{1/2}(\wick) &=& \sin \frac{\wick}{2}\, =\, \sqrt{\frac{1}{2}-\frac{1}{2}\cos\wick} = \nonumber \\
                          & = & \sqrt{\frac{1}{2}-\frac{\gamma}{2p'}(p+\beta E\cos\delta)} . \label{d2}
\end{eqnarray}
\end{subequations}

\noindent
This means that the quark helicity spinors in $S_\cm$ are expressed in terms of Dirac spinors in $\spi$ as 
\begin{subequations}
\begin{eqnarray}
 u(\pvec',\lambda') & = &
   \left ( \begin{array}{c} \sqrt{E+m}\\ \frac{|l^{-1}\pvec|}{\sqrt{E+m}} \end{array} \right )\!
  \chi_{_{\frac{1}{2}}}\!(\mathbf{\hat p'}) d_{1/2,\lambda'}^{1/2}(\wick) +  \nonumber \\
   & + & \!\left ( \begin{array}{c} \sqrt{E+m} \\ \frac{-|l^{-1}\pvec|}{\sqrt{E+m}} \end{array} \right )\! 
  \chi_{_{-{\frac{1}{2}}}}\!(\mathbf{\hat p'}) d_{-1/2,\lambda'}^{1/ 2}(\wick) 
\label{u}
\end{eqnarray} 
and similarly for the antiquark spinors
\begin{eqnarray}
 v(\pvec',\lambda') & = &
  \left ( \begin{array}{c} \frac{-|l^{-1}\pvec|}{\sqrt{E+m}}\\ \sqrt{E+m} \end{array}\right )\!
  \chi_{_{-{\frac{1}{2}}}}\!(\mathbf{\hat p'}) d_{\lambda',1/ 2}^{1/ 2}(\wick) -  \nonumber \\   
   & - & \!\left ( \begin{array}{c} \frac{|l^{-1}\pvec|}{\sqrt{E+m}}\\ \sqrt{E+m} \end{array}\right )\!
  \chi_{_{\frac{1}{2}}}\!(\mathbf{\hat p'}) d_{\lambda',-1/ 2}^{1/ 2}(\wick) .
\label{v}
\end{eqnarray} 
\end{subequations}
To obtain the above expression for antiquark spinors the property $d_{\lambda \lambda'}(-\wick) = d_{\lambda' \lambda}(\wick)$ 
has been used. Note that the spinors now depend on $\tlab$ via the relativistic factor $\gamma$ present in Eqs.~(\ref{d1}) 
and (\ref{d2}) as well as in the Lorentz boost $l^{-1}$.

\section{Relativistic spinor corrections to \textbf{{\em R}2} and \textbf{{\em A}2} diagrams}

Having shown in Sec.~\ref{secthree} how the Lorentz transformation of the spatial part of the intrinsic pion wave function 
alters the annihilation operators, we will now discuss the effect on the operators due to modifications in the spinors of the final 
quarks $i=1',2'$ and antiquarks $i=4',5'$ . The $A2$ and $R2$ spin-matrix elements read schematically  
\begin{eqnarray}
\hspace*{-5mm} M_S & = & \!\!\sum^{+\frac{1}{2}}_{\{\lambda_i\}=-\frac{1}{2}} \Big \langle [s_{1'}s_{5'}]_{S_{\pi}=0} \Big |
       \otimes \Big \langle [s_{2'}s_{4'} ]_{S_{\pi}=0}\Big  |  \times \nonumber  \\ 
 & \times & \hat V_{R2/A2}\,\, \Big | [s_1s_2s_3 ]_{S_{p}=\frac{1}{2}} \Big \rangle\!\otimes 
    \Big | [s_4s_5s_6 ]_{S_{\bar p}=\frac{1}{2}} \Big \rangle  ,
\label{spinmatrix}
\end{eqnarray}
where $\hat V_{R2/A2}$ is either one of the four operators introduced in Eqs.~(\ref{R2operator1}), (\ref{R2operator2}),
(\ref{A2operator1}), and (\ref{A2operator2}). As before, we neglect relativistic corrections to the $\bar pp$
wave functions, hence, only the final quark and antiquark helicity states are Lorentz transformed. Note that the operators 
$\hat V_{R2/A2}$ mentioned above are in the c.m. frame, where they act between initial- and final-state spin wave functions.

In the derivation of the transition operators in Eqs.~(\ref{R2operator1}), (\ref{R2operator2}), (\ref{A2operator1}), and 
(\ref{A2operator2}), only leading order terms in $\pvec$ were kept. Here the same strategy is followed: we make use of 
Eqs.~(\ref{u}) and (\ref{v}) but we do apply these boosts to the Pauli spinors in the pions, each consisting of a quark 
$i$ and antiquark $j$ with spin states $|s_i m_i\rangle$ and $|s_j m_j\rangle$ (with corresponding helicities $\lambda_i$
and $\lambda_j$) and $m_i\neq m_j=\pm \mbox{$\frac{1}{2}$}$:
\begin{displaymath}
  |s_\pi= m_\pi =0 \rangle = \frac{1}{\sqrt{2}}\, \Big ( |s_i m_i \rangle\otimes 
             |s_j m_j \rangle - |s_i m_j \rangle\otimes |s_j m_i \rangle \Big ).
\end{displaymath}
The Pauli spinors in the c.m. frame are expressed in terms of the Pauli spinors in $S_\pi$ by
\begin{subequations}
\begin{equation}
\hspace*{-1mm}\chi^{ }_{\lambda'} (\mathbf{\hat p}') = d_{1/2,\lambda'}^{1/2}(\wick)\chi_{_{\frac{1}{2}}}
  \!(\mathbf{\hat p}') + d_{-\!1/2,\lambda'}^{1/2}(\wick)\chi_{_{-\!\frac{1}{2}}}\!(\mathbf{\hat p}')
\label{pauli1}
\end{equation}
for particles and 
\begin{equation}
\hspace*{-1mm} \chi^{ }_{\lambda'}(\mathbf{\hat p}') = d_{\lambda',1/2}^{1/2}(\wick)\chi_{_{\!\frac{1}{2}}}(\mathbf{\hat p}') +
   d_{\lambda',-1/2}^{1/2}(\wick)\chi_{_{-\!\frac{1}{2}}} (\mathbf{\hat p}') 
\label{pauli2}
\end{equation}
\end{subequations}
for antiparticles.

The Pauli matrices $\bm{\sigma}$ in the transition operators $\hat V_{R2/A2}$ act on the boosted spinors $\chi^i_\lambda$, with 
$i=1',2',4',5'$. The spin-matrix elements pertinent to the $\hat V_{R2/A2}$ operators are thus modified since the rotation matrix 
elements $d_{\pm 1/2,1/2}^{1/2}(\wick)$ in (\ref{d1}) and (\ref{d2}) come into play. We make use of the fact that in the c.m. frame 
the transverse components of the quark momenta are much smaller than the component along the boost direction (from now on we refer only
to the quark but the discussion is identical for the antiquark). We can thus approximate
\begin{eqnarray}
 |l^{-1}(\bm{\beta})\pvec| & = & \sqrt{{p_{_\perp}}^{\!2}+ \gamma^2 (p_{_\|}+\beta E)^2} \nonumber \\
                           & \simeq & \gamma (\pvec\!\cdot {\bf \hat P}_{\!\pi}+\beta E) ,
\label{l1approx}
\end{eqnarray}
which we write as $p' \simeq \gamma (p\cos\delta +\beta E)$. This approximation is valid except for a small region where
$\cos\delta=-\beta E/p$. With this, the following approximation for the matrix elements 
$d_{\pm 1/2,1/2}^{1/2}(\wick)$ holds
\begin{equation}
 d_{\pm1/ 2,1/ 2}^{1/ 2}(\wick) \simeq \sqrt{ \frac{1}{2} \pm \frac{1}{2} \frac{p +\beta E\cos \delta}{p\cos\delta +\beta E}} .
\label{dapprox}
\end{equation}
The numerator and denominator in Eq.~(\ref{dapprox}) are very similar and the matrix elements $d_{\pm 1/2,1/2}^{1/2}(\wick)$ are 
functions of $\beta$ rather than of $\gamma$. 

To illustrate the role of the $\cos\delta$ term in  Eq.(\ref{dapprox}), we consider two cases $\cos \delta = \pm 1$, i.e. for which 
the quark momentum is parallel or anti-parallel to the boost direction $\bm{\beta}$. Recall that for $\cos\delta=\pm 1$, of course, 
$\pvec_{_{\perp}}$ vanishes in which case Eqs.~(\ref{l1approx}) and (\ref{dapprox}) are exact.

The following relationships exist between helicity and momentum orientation of a quark in the c.m. frame:
\begin{eqnarray}
  \mbox{If}\, \cos \delta & = & +1 \, \left \{ 
  \begin{array}{ccc}
     d_{1/2,1/2}^{1/2}(\wick) & = & \! 1 \\
     d_{-1/2,1/2}^{1/2}(\wick) & =  & \! 0                       
  \end{array} \right .  
\label{dlimit1}
\end{eqnarray}
In other words, if the quark momentum in the pion rest frame is aligned with the boost direction, then helicity is conserved under 
the boost. In this case there is no Wick rotation and $\wick=0$. Of course the momentum $\mathbf{p'}$ and boost vector $\bm{\beta}$ 
will also be parallel in the c.m. frame.  

However, if momentum in the pion rest frame is anti-parallel to the boost direction $\bm{\beta}$ 
then helicity in the c.m. frame can flip, as seen from Eq.~(\ref{dlimit2}) and now $\wick=\pi$. 
\begin{eqnarray}
  \mbox{If}\, \cos \delta & = & -1\, \left \{ 
  \begin{array}{ccc}
     d_{1/2,1/2}^{1/2}(\wick) & =  &\! 0 \\
     d_{-1/2,1/2}^{1/2}(\wick) & = & \! 1                       
  \end{array} \right.  
\label{dlimit2}
\end{eqnarray}

The relativistic effect described in Eqs.~(\ref{dlimit1}) and (\ref{dlimit2}) can be interpreted as follows. Assume a quark inside the 
pion with a given helicity and its momentum oriented anti-parallel to the boost direction $\bm{\beta}$. For very large boosts, an 
observer in the c.m. system perceives a reversed quark-momentum. As the spin direction is not changed, this implies a flipping of the 
helicity in the c.m. frame, which confirms that for $\cos \delta = -1$ the matrix element $d_{1/2,1/2}^{1/2}(\wick)$ is zero whereas 
$d_{-1/2,1/2}^{1/2}(\wick)$ is one as seen in Eq.~(\ref{dlimit2}). On the other hand, if the quark momentum in the pion frame is parallel 
with $\bm{\beta}$, then helicity will not flip, as the quark momentum in the c.m. is still seen in boost direction and $d_{-1/2,1/2}^{1/2}(\wick)$ 
must be zero in agreement with Eq.~(\ref{dlimit1}). Hence, helicities can flip from one reference frame to the other depending on the initial 
orientation of the quark momentum with respect to the boost direction.

For intermediate angles $-1\leq \cos\delta \leq +1$, Eqs.~(\ref{pauli1}) and (\ref{pauli2}) describe the degree of helicity flip
due to the change of reference frame.
Since the rotation matrices are proportional to $\beta$, the relativistic effects from the spinors tend to be much smaller than those 
from the spatial components
discussed previously in Section III.

\section{Conclusion}

In this paper, the effects of Lorentz transformations on intrinsic pion wave functions in the reaction $\pppi$ are investigated. 
This is done within the framework of a constituent quark model used previously \cite{kloet2,kloet1,elbennich1} to describe 
$d\sigma/d\Omega$ and $A_{0n}$ data of this reaction. The Lorentz transformations are effected on both the spatial part as well as 
the quark spinor components in the intrinsic pion wave functions. We find that the coordinate-space wave function is strongly contracted 
along the boost direction and that the spatial part of the annihilation amplitudes depends on $\gamma$ and $\gamma^2$. The spinors 
are modified by Wick rotations and the relativistic effects due to spin are therefore proportional to $\beta$ rather than to 
the boost factor $\gamma$. 

In addition to the non-planar quark rearrangement diagrams R2, we also compute the transition operators originating from the planar 
$A2$ diagrams for pure $\s$ as well as a mixture of $\s$ and $\p$ annihilation mechanisms. 

The non-relativistic annihilation operator for the non-planar $R2$ diagrams was given in Ref.~\cite{kloet2}. In this paper we obtain new 
terms in the operator that affect its range and introduce additional angular dependence. The dependence on the c.m. energy $\sqrt{s}$ is 
now manifest via the relativistic boost factor $\gamma$. Most of the new relativistic factors are considerably larger than the non-relativistic 
ones. Therefore, the geometry of the annihilation amplitudes is drastically modified. Nonetheless the selection rules described in Ref.~\cite{kloet2} 
are preserved. The $A2$ amplitudes are simpler than the $R2$ amplitudes and they depend only on the relative distance of the proton and 
antiproton. 

We will show \cite{elbennich2} that the modifications in the annihilation amplitudes, due to relativistic considerations, strongly improve 
the description of the cross section and analyzing power as measured at LEAR \cite{hasan}. Therefore with a relativistic treatment one obtains 
a description of the reaction $\pppi$ similar or better than with the ad hoc increase of the particle radii which was the approach in 
Ref.~\cite{kloet2}.

\begin{acknowledgments}
B.E. would like to thank Beno\^{\i}t Loiseau for helpful discussions and the Laboratoire de Physique Nucl\'eaire et de
Hautes Energies (Groupe Th\'eorie) at the Universit\'e Pierre et Marie Curie for its hospitality and support. Part of 
this work was carried out at the Institute for Nuclear Theory, University of Washington, to whom he is grateful for a 
welcoming atmosphere and support during his stay. W.M.K acknowledges a very stimulating discussion with Stefan Scherer. 
The effects of Lorentz transformations of spinors were investigated due to stimulating comments by Ronald Gilman.
\end{acknowledgments}

\end{document}